\documentclass[fleqn,twoside,twocolumn,nofootinbib,showkeys]{revtex4} 
\usepackage[sec,nocpr]{ujp} 
\usepackage{graphicx}
\usepackage{dcolumn}
\usepackage{bm}
\begin{document}
\title[Strangeness enhancement  at  the hadronic chemical freeze-out]
{STRANGENESS ENHANCEMENT AT THE HADRONIC CHEMICAL FREEZE-OUT}%
\author{V.V.~Sagun}
\affiliation{\bitp}
\address{\bitpaddr}
\email{v_sagun@ukr.net}
\author{D.R. Oliinychenko}
\affiliation{\bitp}
\address{\bitpaddr}
\email{dimafopf@gmail.com}
\affiliation{FIAS,Goethe-University, }
\address{Ruth-Moufang Str. 1, 60438 Frankfurt upon Main, Germany}
\author{K.A.~Bugaev}
\affiliation{\bitp}
\address{\bitpaddr}
\email{bugaev@th.physik.uni-frankfurt.de}
\author{J. Cleymans}
\affiliation{Department of Physics, University of Cape Town}
\address{Rondebosch 7701, South Africa}
\email{jean.cleymans@uct.ac.za}

\author{A.I. Ivanytskyi}
\affiliation{\bitp}
\address{\bitpaddr}
\email{a_iv_@ukr.net}

\author{I.N. Mishustin}
\affiliation{FIAS, Goethe-University, }
\address{Ruth-Moufang Str. 1, 60438 Frankfurt upon Main, Germany}
\email{mishustin@fias.uni-frankfurt.de}
\affiliation{Kurchatov Institute, Russian Research Center}
\address{Kurchatov Sqr., Moscow, 123182,  Russia}

\author{E.G. Nikonov}
\affiliation{Laboratory for Information Technologies, JINR}
\address{141980 Dubna, Russia}
\email{e.nikonov@jinr.ru}

 \udk{539.12}  \razd{\seci}

\setcounter{page}{1}%
\maketitle

\begin{abstract}
The chemical freeze-out of hadrons created in the high energy nuclear collisions is studied within the realistic version of  the hadron resonance gas model.  The chemical  non-equilibrium of  strange particles is accounted  via the usual $\gamma_{s}$ factor which gives  us an opportunity  to perform a high quality fit with $\chi^2/dof \simeq  63.5/55 \simeq$ 1.15 of the hadronic multiplicity ratios measured from the low AGS to the highest RHIC energies. In contrast to previous findings,  at low energies we observe the strangeness enhancement instead of  a suppression.
In addition, the performed $\gamma_{s}$ fit allows us  to achieve the  highest quality of  the Strangeness Horn description with  $\chi^2/dof=3.3/14$. For the first time the top point of the  Strangeness Horn is perfectly reproduced, which makes our theoretical horn as sharp as an experimental one.
However, the  $\gamma_{s}$ fit approach does not sizably  improve the description of the multi-strange baryons and antibaryons. Therefore,  an apparent deviation of  multi-strange baryons and antibaryons  from chemical equilibrium requires further explanation.
\end{abstract}


\section{Introduction} \label{Introduction}

The   hadron yields  measured  in  heavy ion collisions are traditionally analyzed by the
Hadron Resonance Gas Model (HRGM) \cite{Thermal_model_review,KABAndronic:05,Becattini:gammaHIC,KABugaev:Horn2013}. Its main assumption is an existence of the thermal equilibrium in the system under consideration, which is strongly supported by an excellent coincidence of experimental and theoretical yields of the hadrons built up from $u$ and $d$ quarks. Using the temperature $T$, the baryonic  chemical potential $\mu_B$ and isospin third component chemical potential
$ \mu_{I3}$ the HRGM  allows one  to describe the hadronic multiplicities at the moment 
of chemical freeze-out (FO), the moment at which all inelastic reactions  cease to exist. 
However, the HRGM has some traditional  problems to describe  the strange hadrons. Thus,  within the standard
HRGM formulation with a single hard-core repulsion radius for all particles 
the energy dependences of ${K^+}/{\pi^+}$ and $\Lambda/\pi^-$ ratios were never  satisfactorily  reproduced  \cite{Thermal_model_review,KABAndronic:05,Becattini:gammaHIC}. However, 
the non-monotonic energy dependence of ${K^+}/{\pi^+}$ ratio,  known  as  the Strangeness Horn,  is of special interest
because it may  serve as a signal of the onset of deconfinement. 
In order to account for   a deviation from the chemical equilibrium of strange hadrons    the  factor $\gamma_{s}$ was 
introduced \cite{Rafelski:gamma}. It is  used to describe  the undersaturated (oversaturated)   densities  $\gamma_{s}<1$ ($\gamma_{s}>1$)   of each  strange charge.   Formally,  the strange fugacity, associated with each (anti)strange charge,  
is simply  multiplied  by  the $\gamma_{s}$ factor. Although 
the $\gamma_{s}$ factor  plays an important role in the  analysis of  the data  measured  in the collisions of elementary particles \cite{Becattini:gammaHIC} and in  the nucleus-nucleus collisions  \cite{Becattini:gammaHIC, PBM:gamma},   the problem of its justification remains   unsolved. Also the results on the $\gamma_{s}$ values obtained within different thermal models 
are controversial.  For instance,   a strong suppression of the strange charge in nucleus-nucleus collisions  was reported in Ref. \cite{Becattini:gammaHIC},
while the results of Ref. \cite{PBM:gamma} are consistent with $\gamma_{s}=1$. 
In order to resolve the latter  problem,  we apply the most successful version of the  HRGM with multicomponent hard-core 
repulsion \cite{KABugaev:Horn2013, MultiComp:08, MultiComp:12,MultiComp:13b,BugaevEPL:13} to 
describe   111 independent hadron yield ratios  measured at  14 values of the center of mass collision energy $\sqrt{s_{NN}}$ in the interval from 2.7 GeV to 200 GeV.

The work  is organized as follows. In the next section  we give the theoretical basis of the present  model.
In Section 3 we present the data descriptions   and  discuss  them. Section 4 is  devoted to our  conclusions.

\section{Hardon Resonance Gas Model} \label{Model}

We employ  a multicomponent HRGM,  which currently provides the best description of 
the observed hadronic multiplicities. It is the model developed  in \cite{KABugaev:Horn2013, MultiComp:08, MultiComp:12,MultiComp:13b,BugaevEPL:13}. The hadron interaction is taken into account via the  hard-core repulsion whose  radii have  different  values for pions $R_{\pi}$, kaons $R_K$, other mesons $R_m$ and baryons $R_b$. The best  global fit  of all hadronic multiplicities was found    for $R_b$ = 0.2 fm, $R_m$ = 0.4 fm, $R_{\pi}$ = 0.1 fm, $R_K$ = 0.38 fm   \cite{KABugaev:Horn2013}. The main equations of this  HRGM  are listed below, but more details  can be found in \cite{KABugaev:Horn2013,MultiComp:08, MultiComp:12,MultiComp:13b,BugaevEPL:13}.

Let us consider  the Boltzmann gas of $N$ hadron species in a volume $V$ that has  the temperature $T$, the baryonic chemical potential $\mu_B$, the  strange chemical potential $\mu_S$ and the chemical potential of the isospin third component $\mu_{I3}$. The system  pressure $p$ and the $K$-th charge density $n^K_i$ ($K\in\{B,S, I3\}$) of the  i-th hadron sort are given by the expressions  
\begin{eqnarray}\label{EqI}
%
%
%
\frac{p}{T} =  \sum_{i=1}^N \xi_i \,, ~~n^K_i =\frac{ Q_i^K{\xi_i}}{\textstyle  1+\frac{\xi^T {\cal B}\xi}{\sum\limits_{j=1}^N \xi_j}}, ~~\xi  = \left(
\begin{array}{c}
 \xi_1 \\
 \xi_2 \\
 ... \\
 \xi_N
\end{array}
\right), 
\end{eqnarray}
where $\cal B$ denotes a symmetric  matrix of the second  virial coefficients with the elements $b_{ij} = \frac{2\pi}{3}(R_i+R_j)^3$ and 
 the variables $\xi_i$ are the solutions of the following system
\begin{eqnarray}\label{EqII}
&&\hspace*{-4mm}\xi_i =\phi_i (T)\,   \exp\Biggl[ \frac{\mu_i}{T} - {\textstyle \sum\limits_{j=1}^N} 2\xi_j b_{ij}+{\xi^T{\cal B}\xi} {\textstyle \left[ \sum\limits_{j=1}^N\xi_j\right]^{-1}} \Biggr] \,, \quad \quad \\
&&\hspace*{-4mm}\phi_i (T)  = \frac{g_i}{(2\pi)^3}\int \exp\left(-\frac{\sqrt{k^2+m_i^2}}{T} \right)d^3k  \,.
\label{EqIII}
\end{eqnarray}
Here   the full chemical potential of the $i$-th hadron sort $\mu_i \equiv Q_i^B \mu_B + Q_i^S \mu_S + Q_i^{I3} \mu_{I3}$ is expressed in terms of the corresponding charges $Q_i^K$  and their  chemical potentials,  $ \phi_i (T) $ denotes 
the thermal particle  density of  the $i$-th hadron sort of mass $m_i$ and degeneracy $g_i$, and  $\xi^T$  denotes  the row of  variables $\xi_i$.  
For each collision energy 
the fitting 
parameters are temperature $T$, baryonic chemical potential $\mu_B$ and the chemical potential of the third projection of  isospin  $\mu_{I3}$, whereas the strange chemical potential  $\mu_S$ is found from the   condition of  vanishing strangeness.

In order to  account for the possible  strangeness non-equilibration we introduce the  $\gamma_s$ factor in a conventional way \cite{Rafelski:gamma} by replacing $\phi_i$ in Eqs. (\ref{EqII}) and (\ref{EqIII})  as
\begin{eqnarray} \label{EqIV}
\phi_i(T) \to \phi_i(T) \gamma_s^{s_i} \,,
\end{eqnarray}
where $s_i$ is a number of strange valence quarks plus number of strange  valence anti-quarks.

Width correction is taken into account by averaging all expressions containing mass by Breit-Wigner distribution having a  threshold. As a result, the modified thermal particle density of $i$-th hadron  sort  acquires  the form
\begin{eqnarray}
\label{EqV}
&&\int \exp\left(-\frac{\sqrt{k^2+m_i^2}}{T} \right)d^3k \rightarrow \nonumber \\
&& \frac{\int^{\infty}_{M_{0}} \frac{dx}{(x-m_{i})^{2}+\Gamma^{2}_{i}/4} \int \exp\left(-\frac{\sqrt{k^2+x^2}}{T} \right)d^3k }{\int^{\infty}_{M_{0}} \frac{dx}{(x-m_{i})^{2}+\Gamma^{2}_{i}/4}} \,. 
\end{eqnarray}
Here $m_i$ denotes the  mean mass of  hadron and $M_0$ stands for the threshold in the dominant decay channel. The main advantages of this approximation  are a simplicity of  its realization and a clear way to account for the finite  width of hadrons. The effect of the  vanishing resonances width or the Gaussian resonance width parameterization on the chemical FO parameters is discussed  in  \cite{Bugaev:1312a}.

The effect of resonance decay $Y \to X$ on the final hadronic multiplicity is taken into account as $n^{fin}(X) = \sum_Y BR(Y \to X) n^{th}(Y)$, where $BR(X \to X)$ = 1 for the sake of convenience. 
The masses, the  widths and the strong decay branchings of all experimentally known hadrons  were  taken from the particle tables  used  by  the  thermodynamic code THERMUS \cite{THERMUS}.

\section{Results} \label{Results}

{\bf Data sets  and fit procedure}. In this work  we use the data set which is  identical to Ref. \cite{BugaevEPL:13}. At the AGS energies ($\sqrt{s_{NN}}=2.7-4.9$ AGeV or $E_{lab}=2-10.7$ AGeV) the data are available with a  good energy resolution above 2 AGeV. However, for the beam energies 2, 4, 6 and 8 AGeV only a few data points are available. They correspond  to the yields for pions \cite{AGS_pi1, AGS_pi2}, for protons \cite{AGS_p1,AGS_p2}, for kaons \cite{AGS_pi2} (expect for 2 AGeV). The integrated over $4\pi$ data are also available for $\Lambda$ hyperons \cite{AGS_L} and for $\Xi^-$ hyperons (for 6 AGeV only) \cite{AGS_Kas}. 
\begin{figure}[t]
\centerline{\includegraphics[width=7.cm, height=6.5cm]{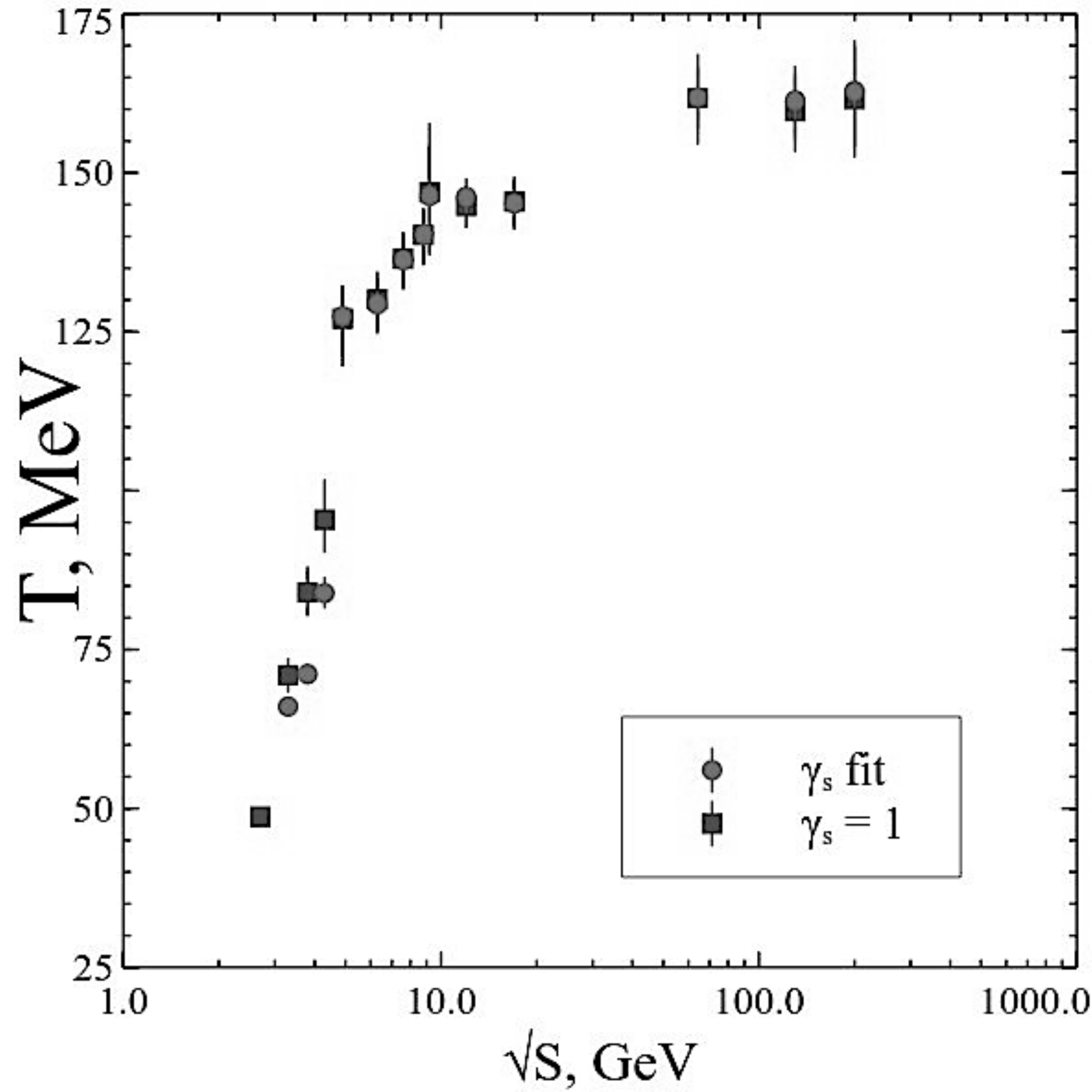}}
\centerline{\includegraphics[width=7cm, height=6.5cm]{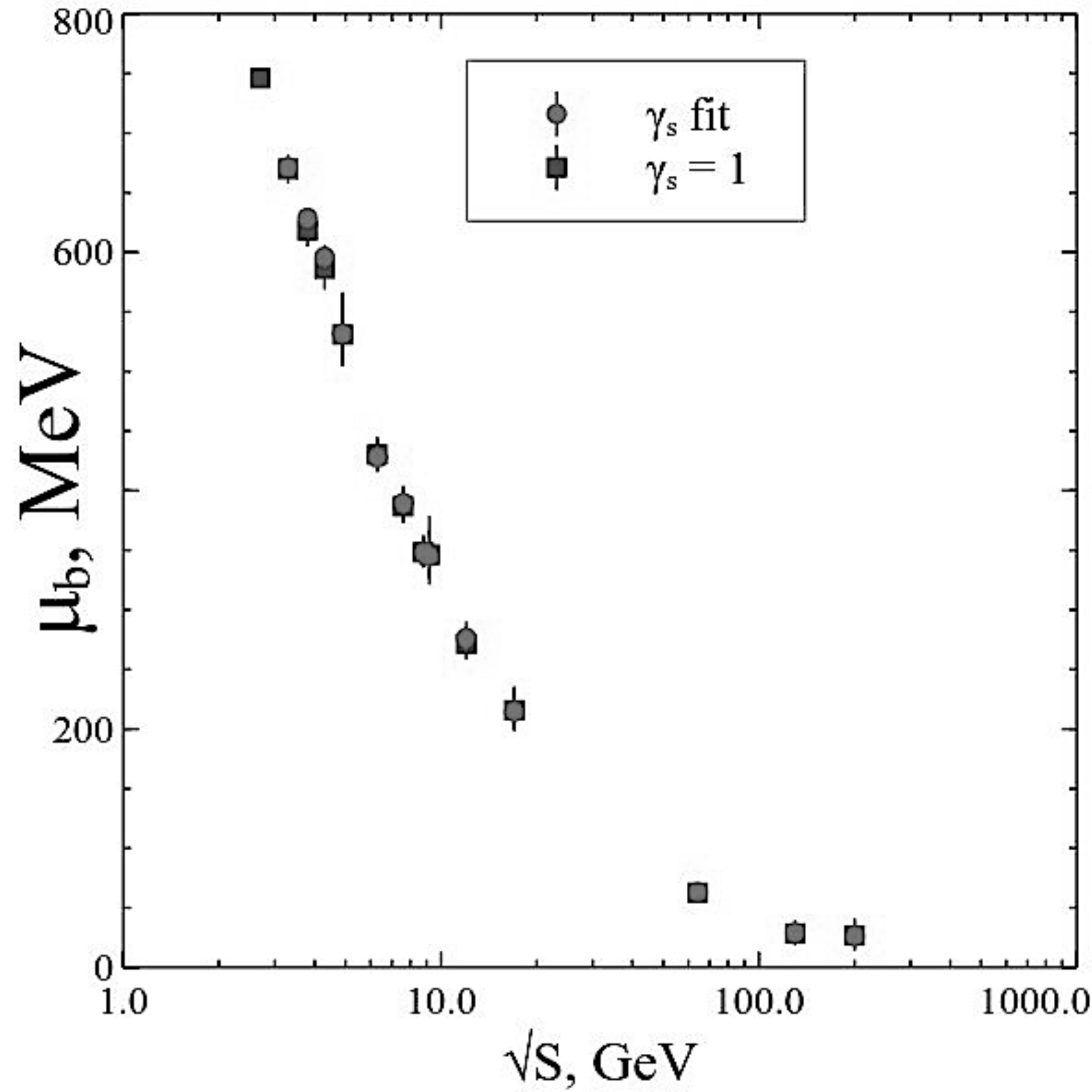}}
\vspace*{-3mm}
\caption{
Comparison of the  parameters behavior for the  $\gamma_s$  fit and for the  chemical FO model  with $\gamma_s$ = 1. 
Upper panel: the chemical FO temperature $T$. 
Lower panel: the chemical FO  baryo-chemical potential $\mu_B$.
}
\label{fig_SagunI}
\end{figure}
However, as it  was argued in Ref. \cite{KABAndronic:05}, the data for $\Lambda$ and $\Xi^-$ should be recalculated for midrapidity. Therefore, instead of raw experimental data we used the corrected values from \cite{KABAndronic:05}. Further we analyzed  the data set at the highest AGS energy ($\sqrt{s_{NN}}=4.9$ AGeV or $E_{lab}=10.7$ AGeV). Similarly to \cite{KABugaev:Horn2013}, here  we analyzed  only  the  NA49  mid-rapidity data   \cite{KABNA49:17a,KABNA49:17b,KABNA49:17Ha,KABNA49:17Hb,KABNA49:17Hc,KABNA49:17phi}
as the most difficult ones  to be reproduced. 
Since  the RHIC high energy  data of different collaborations agree  well with each other, we  analyzed  the STAR results  for  $\sqrt{s_{NN}}= 9.2$ GeV \cite{KABstar:9.2}, $\sqrt{s_{NN}}= 62.4$ GeV \cite{KABstar:62a}, $\sqrt{s_{NN}}= 130$ GeV \cite{KABstar:130a,KABstar:130b,KABstar:130c,KABstar:200a} and  200 GeV \cite{KABstar:200a,KABstar:200b,KABstar:200c}.

The fit criterion  is a minimization  of $\chi^2=\sum_i\frac{(r_i^{ther}-r_i^{exp})^2}{\sigma^2_i}$, where $r_i^{theor}$ and $r_i^{exp}$ are, respectively,  theoretical and experimental  values of particle yield ratios, $\sigma_i$ stands for the corresponding experimental error and a  summation is performed over all available experimental points.

\begin{figure}[t]
\includegraphics[width=7cm, height=6.5cm]{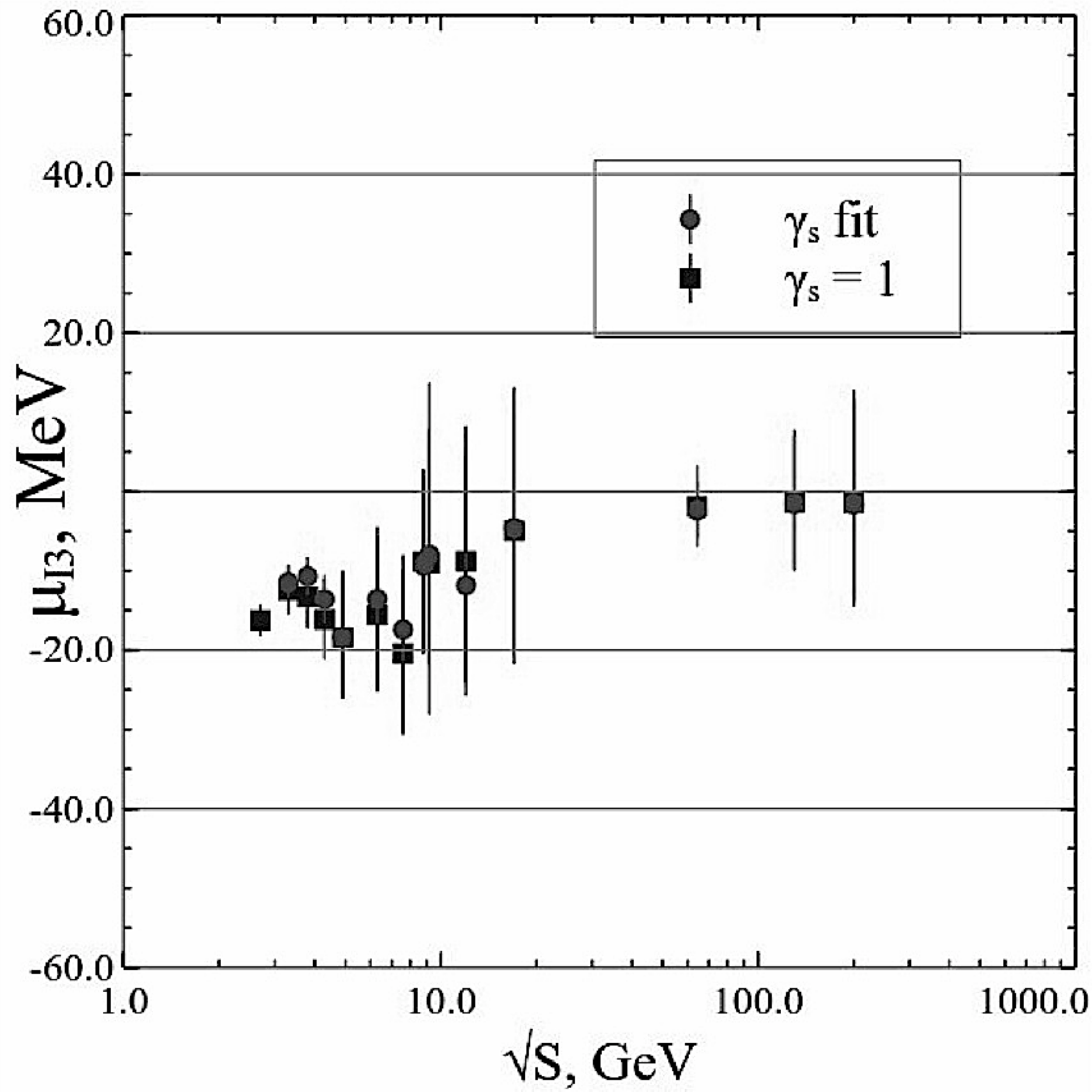}\\
\vspace*{0.cm}
\includegraphics[width=7cm, height=6.5cm]{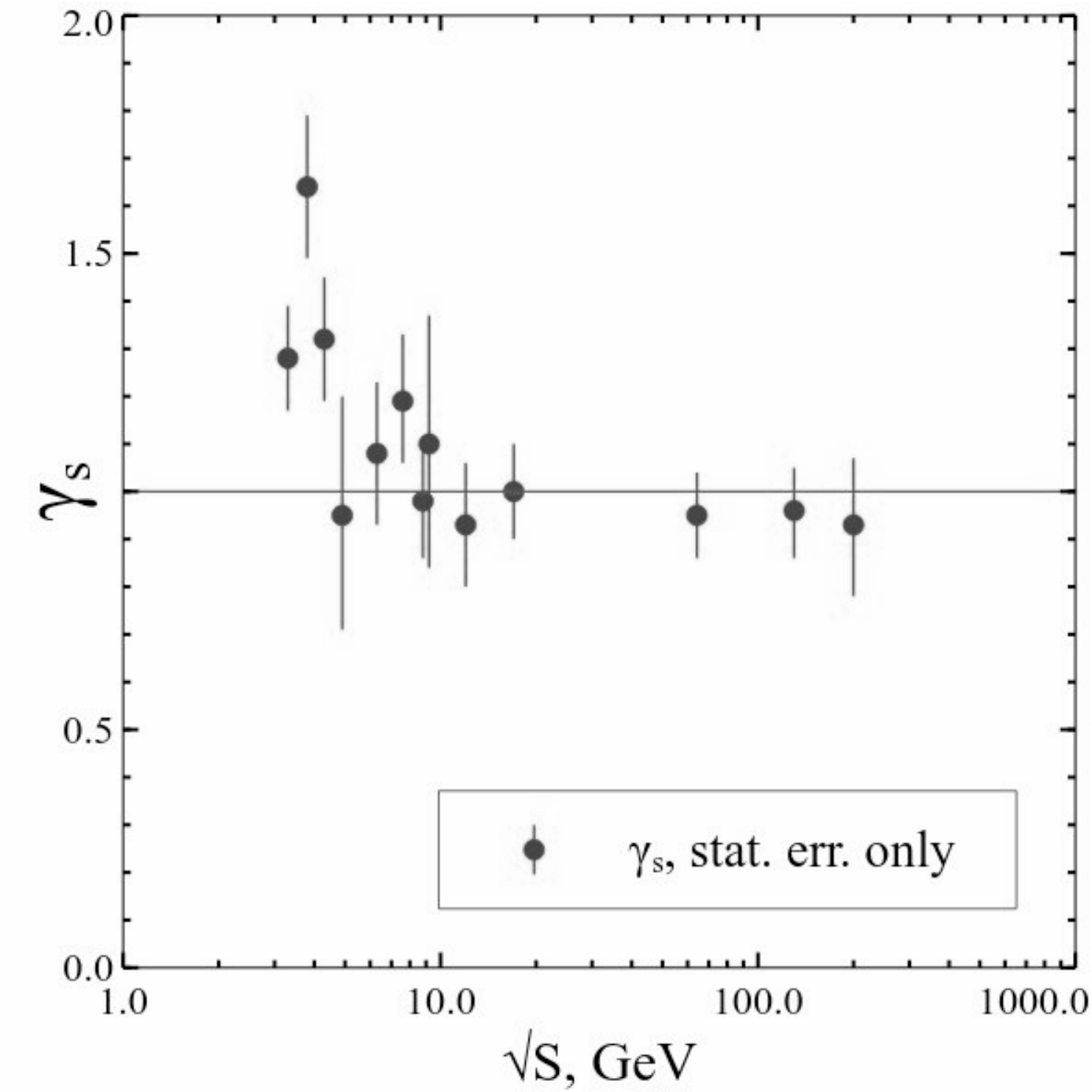}
\vspace*{-3mm}
\caption{ Same as in Fig. (\ref{fig_SagunI}), but for  the chemical potential of the third projection of isospin $\mu_{I3}$ (upper panel)  and   the strangeness enhancement factor $\gamma_s$  (lower panel) .
}
\label{fig_SagunIb}
\end{figure}

\begin{figure}
\includegraphics[width=6.3cm, height=6.3cm]{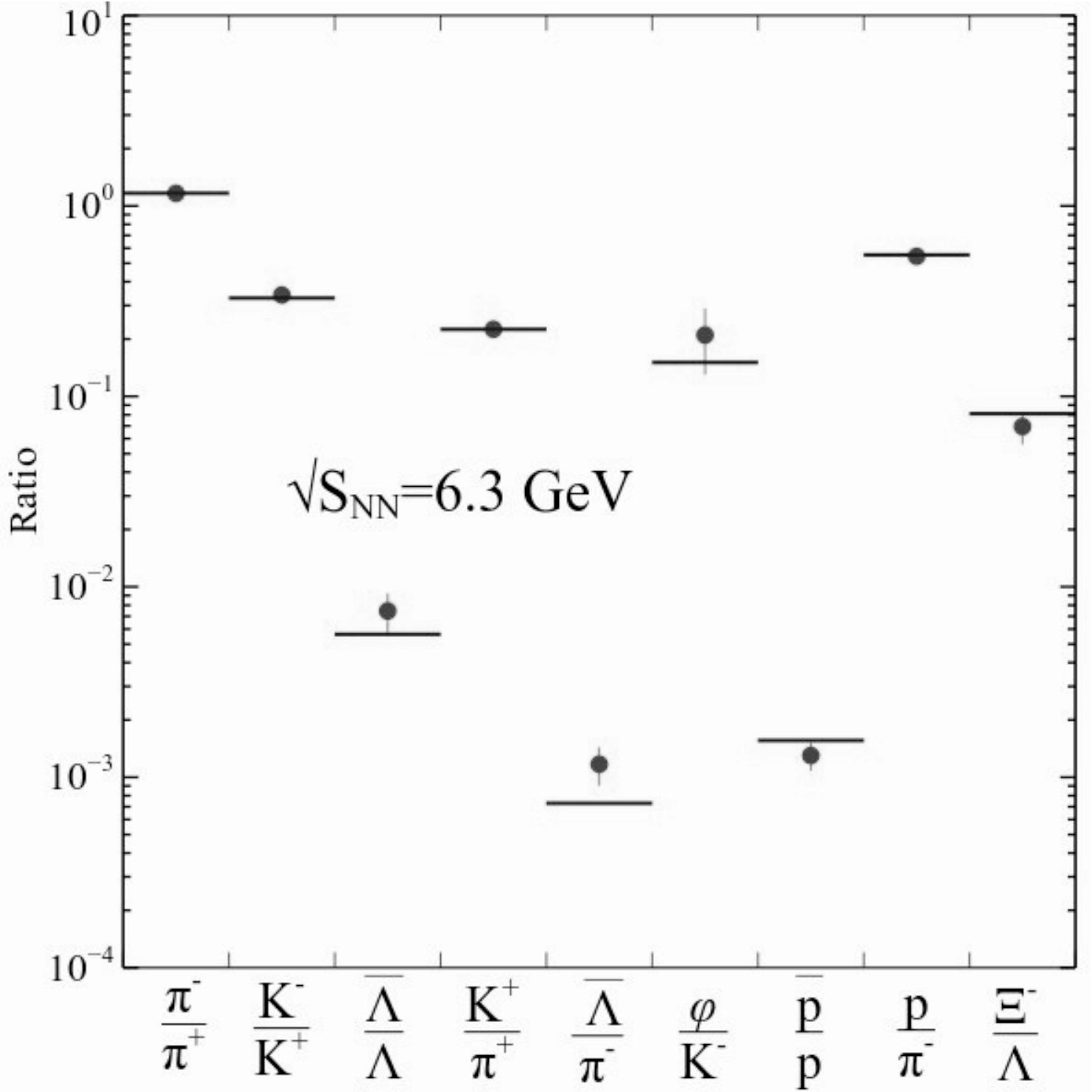}\\
\includegraphics[width=6.3cm, height=6.3cm]{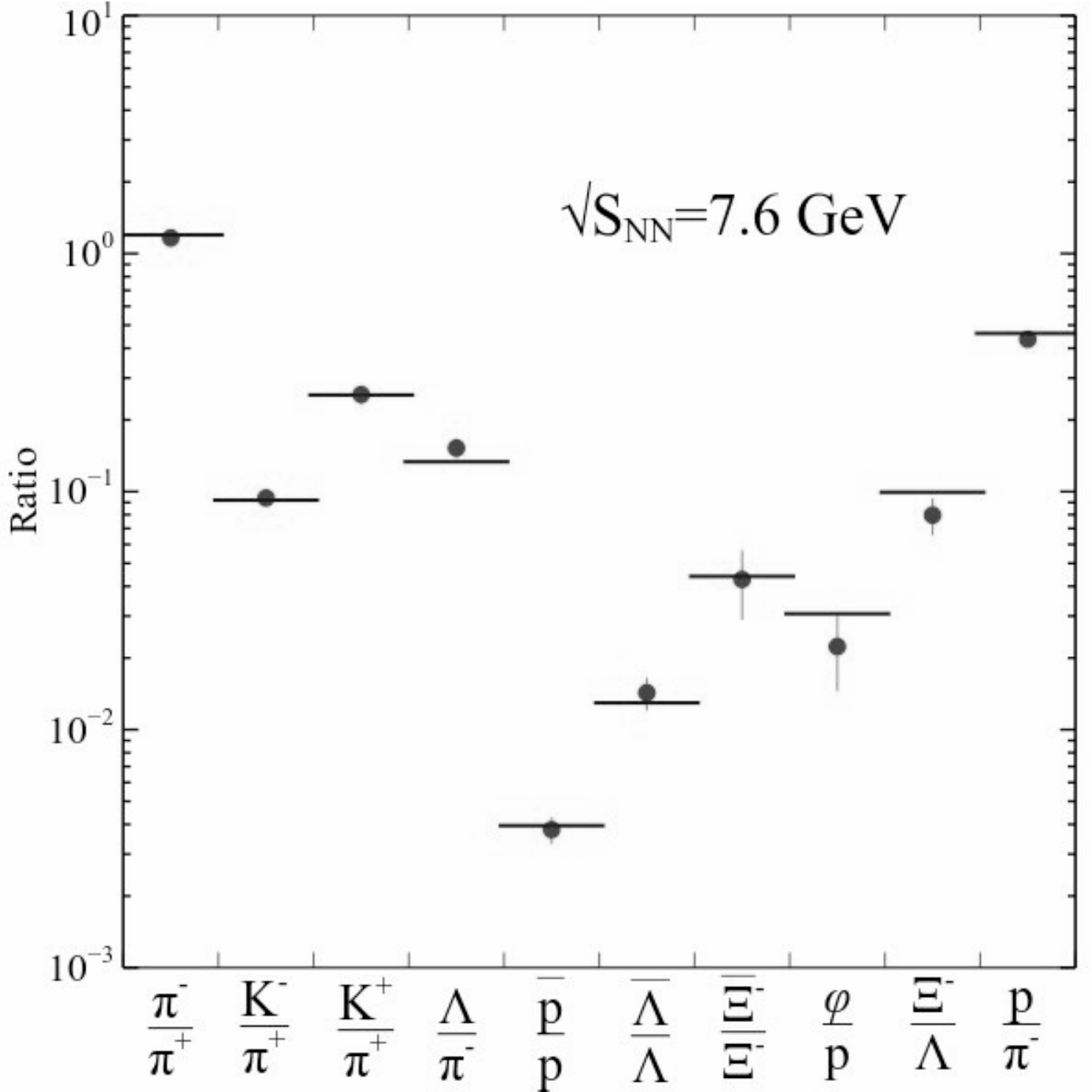}\\
\includegraphics[width=6.3cm, height=6.3cm]{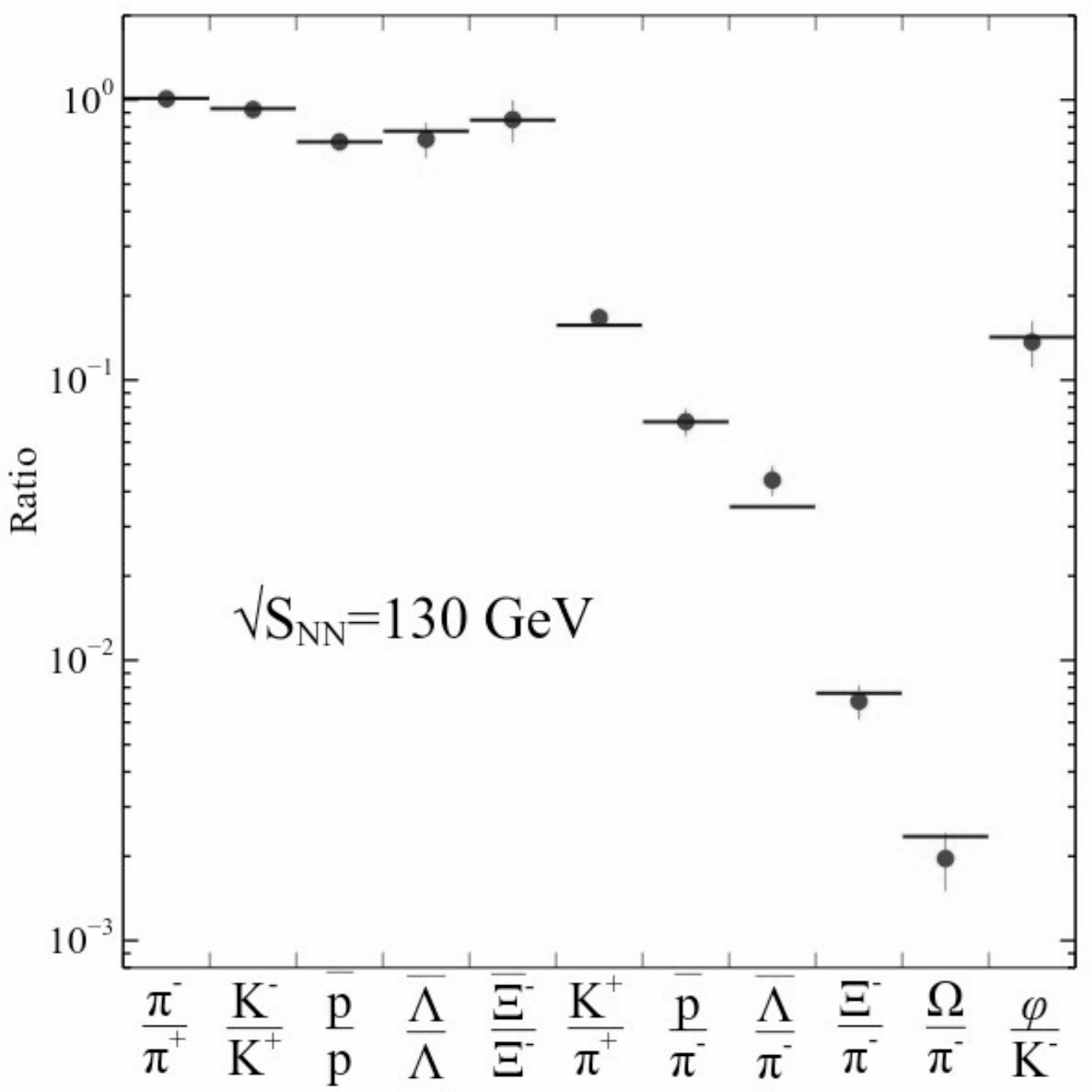}
\vspace*{-3mm}
\caption{The examples of the particle  yield ratios description obtained for the $\gamma_s$ fit. The dots denotes the experimental values,  while the lines show the result of fit. 
The symbols on OX axis demonstrate the particle ratios. 
Upper panel:  $\sqrt{S_{NN}}$ = 6.3 GeV, T= 129 MeV, $\mu_b$ = 427 MeV. 
Middle panel: $\sqrt{S_{NN}}$ = 7.6 GeV, T= 136 MeV, $\mu_b$ = 389 MeV. 
Lower panel: $\sqrt{S_{NN}}$ = 130 GeV, T=161 MeV, $\mu_b$ = 29 MeV.
}
\label{fig_SagunII}
\end{figure}

\begin{figure}
\includegraphics[width=6.3cm, height=6.3cm]{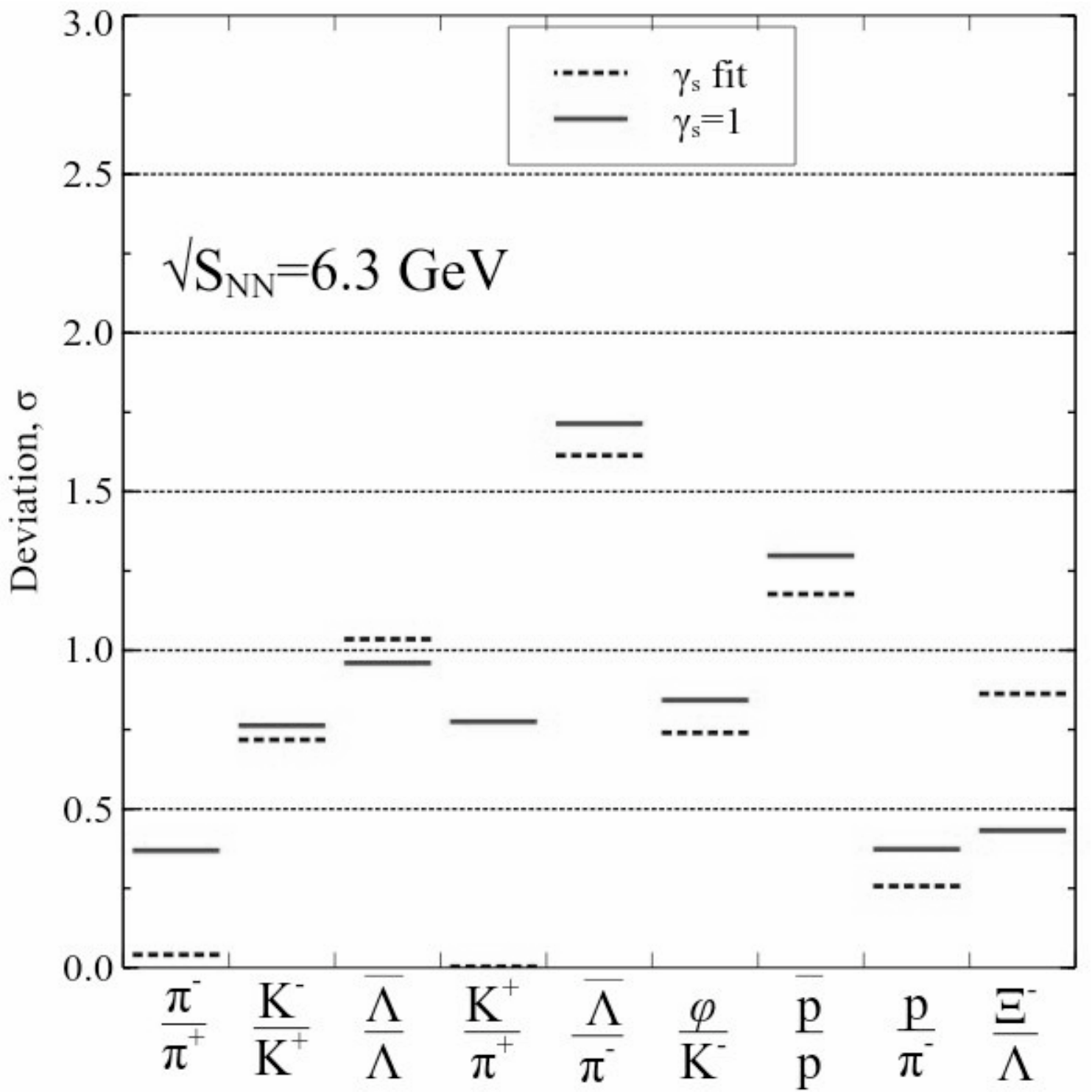}\\
\includegraphics[width=6.3cm, height=6.3cm]{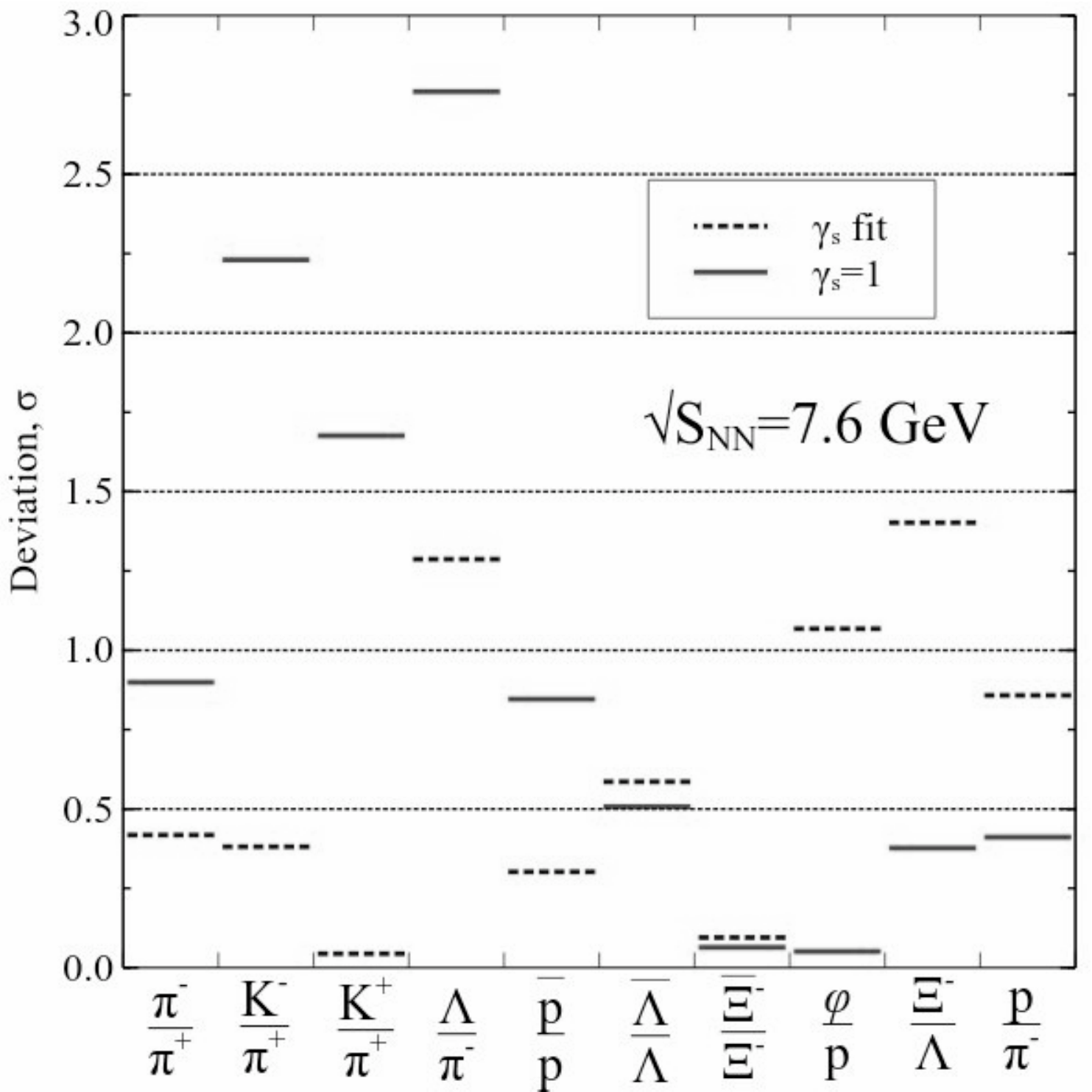}\\
\includegraphics[width=6.3cm, height=6.3cm]{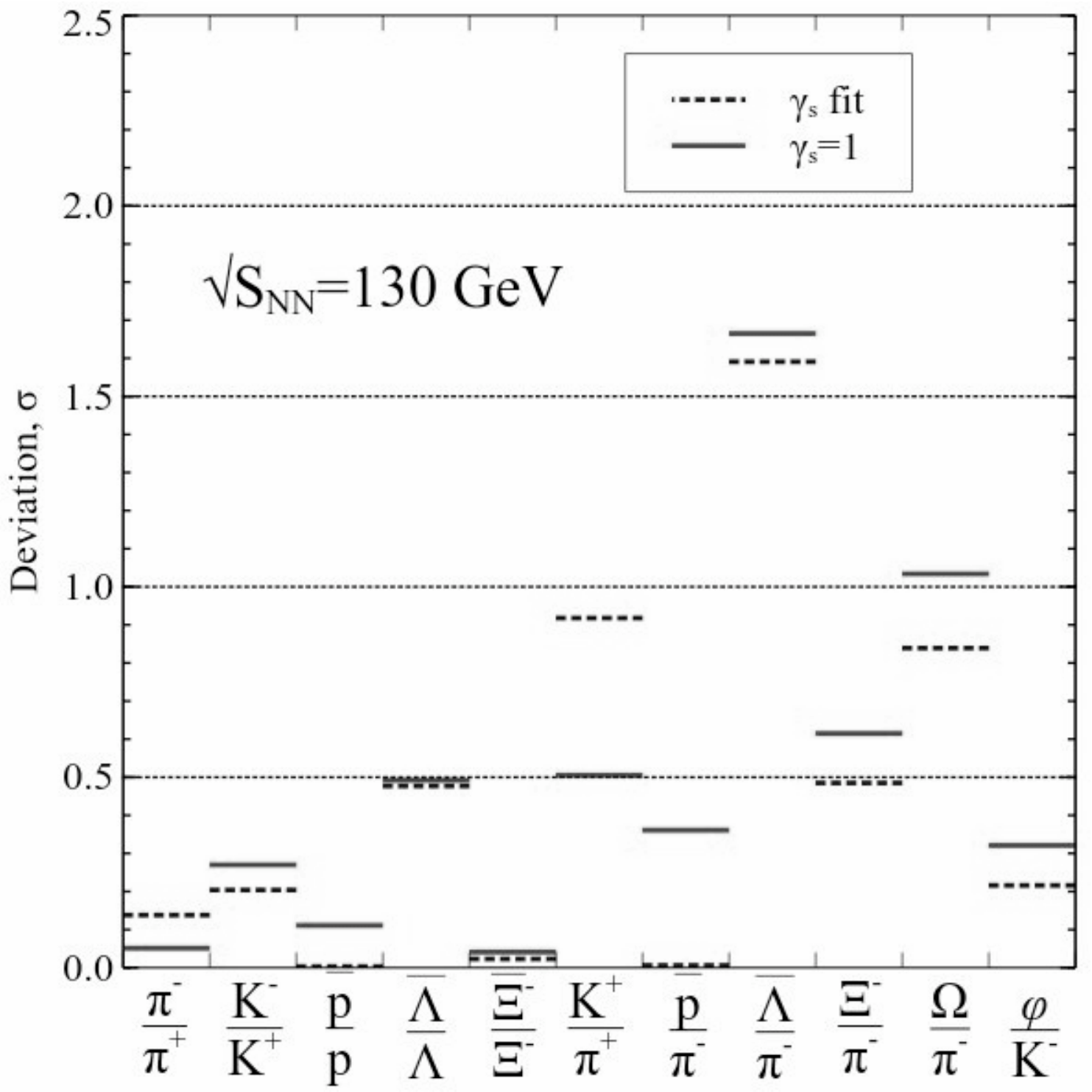}
\vspace*{-3mm}
\caption{
Relative deviation of theoretical description of ratios from experimental value in units of experimental error $\sigma$. The symbols on OX axis demonstrate the particle ratios. OY axis shows $\frac{|r^{theor} - r^{exp}|}{\sigma^{exp}}$, i.e. the modulus of  relative deviation for $\sqrt{s_{NN}}$ = 6.3, 7.6 and 130 GeV. 
  Solid lines correspond to the model of chemical  FO with  $\gamma_s =1$, while the dashed lines correspond to the model with the  $\gamma_s$ fit.
}
\label{fig_SagunIII}
\end{figure}

{\bf The $\gamma_s$ fit}. To improve the description of the strange hadrons  and to 
investigate the role of  their chemical non-equilibrium within the multicomponent 
HRGM we  consider  the $\gamma_s$ factor as a fitting parameter for each value of collision energy.
In our analysis we pay a special  attention  to the  ${K^+}/{\pi^+}$ ratio, because it is  usually considered as the most problematic one for the HRGM.

\begin{figure}[t]
\includegraphics[width=77mm, height=70mm]{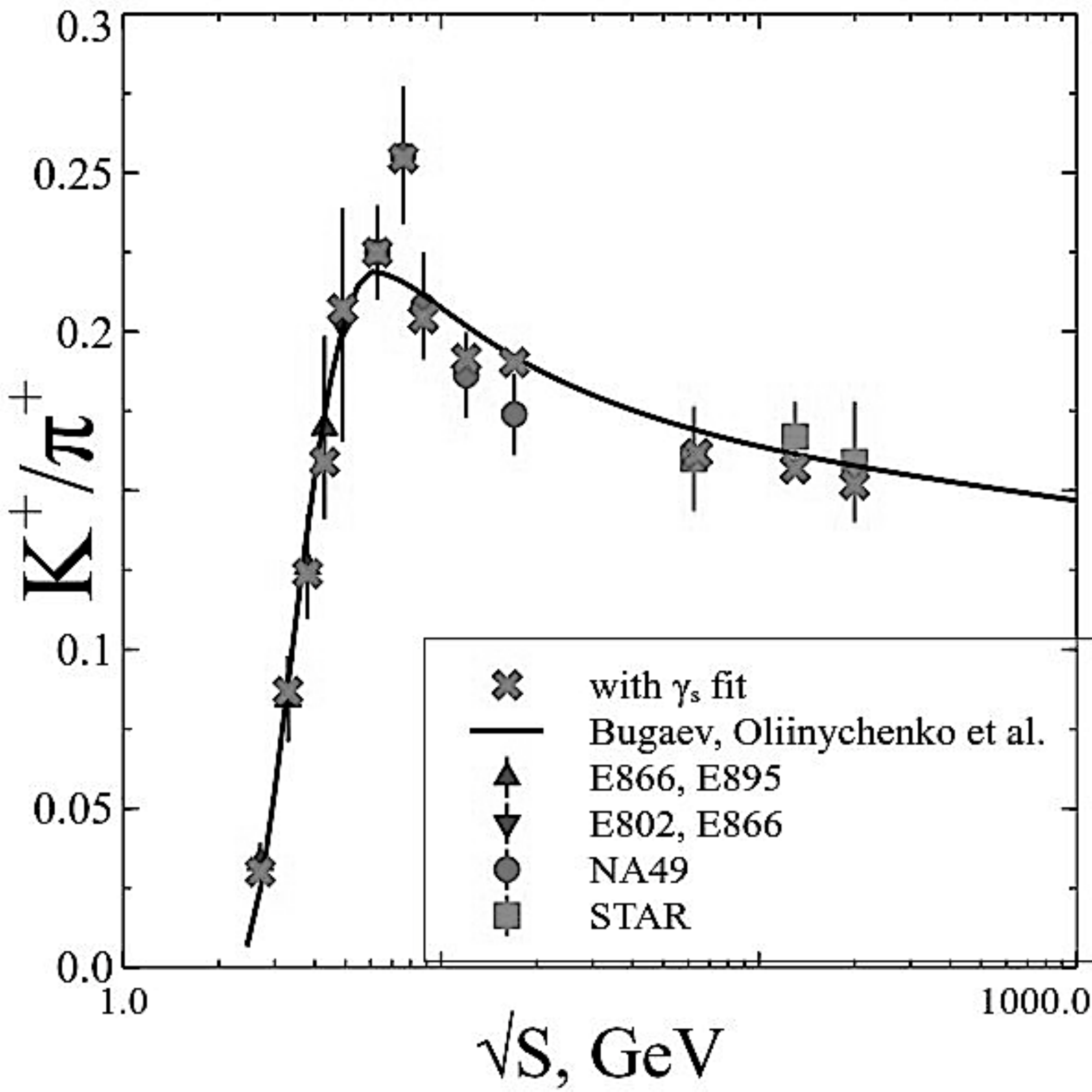}\\
\includegraphics[width=77mm, height=70mm]{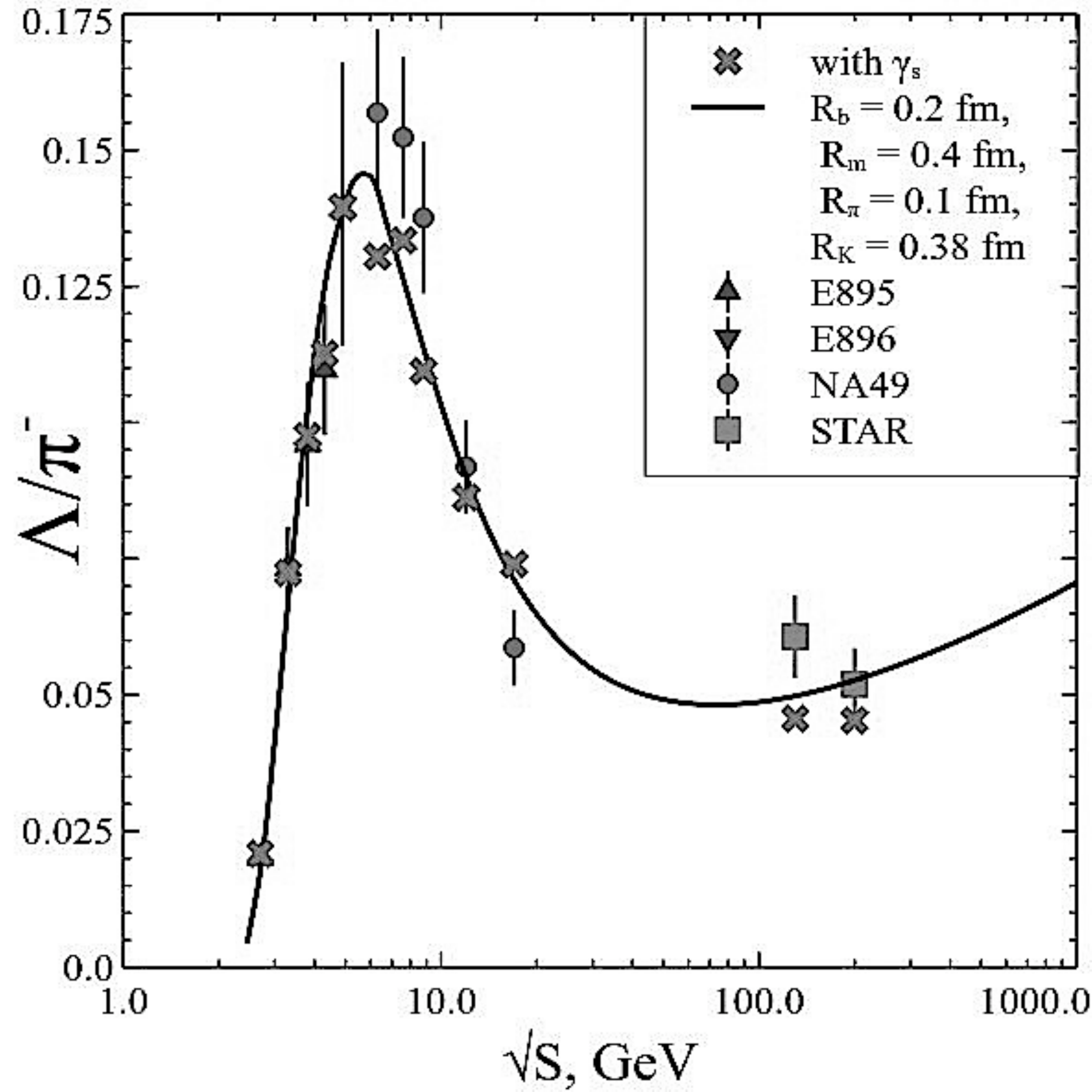}\\
\vspace*{-3mm}
\caption{
Description of ${K^+}/{\pi^+}$ ratio (upper plot) and $\Lambda/\pi^-$ ratio (lower plot). Solid lines show the results  of \cite{KABugaev:Horn2013} for the HRGM with $\gamma_s=1$. Crosses stand for the case of the  $\gamma_s$ fit.
}
\label{fig_SagunIV}
\end{figure}

For  14 collision energies
$\sqrt{s_{NN}} = $  2.7, 3.3, 3.8, 4.3, 4.9, 6.3, 7.6, 8.8, 9.2, 12, 17, 62.4, 130, 200 GeV the resulting 
fit quality $\chi^2/dof$ = 63.4/55 = 1.15 became slightly better compared to the  result $\chi^2/dof$ = 80.5/69 = 1.16 found for the chemical FO model   with $\gamma_s=1$, although  the  value of $\chi^2$ itself, not divided by number of degrees of freedom,  has improved notably. 
As one can see from Figs.~\ref{fig_SagunI} and \ref{fig_SagunIb}, the temperature, the baryo-chemical potential and the chemical potential of the isospin third projection obtained for the $\gamma_s$  fit demonstrate almost the same behavior as for the case 
of  the chemical FO model with $\gamma_s$ = 1. The most remarkable  behavior is demonstrated by  the $\gamma_s(\sqrt{s_{NN}})$ function (see the lower panel of  \ref{fig_SagunIb}). 
In contrast to the earlier results \cite{Becattini:gammaHIC} we found not a strangeness suppression, but a sizable enhancement ($\gamma_s$>1) at the  energies below $\sqrt{s_{NN}} = 8.8$ GeV, while at higher energies our 
results $\gamma_s \simeq 1$ are consistent with the findings of Ref. \cite{PBM:gamma}.
We have to stress, that our results on the  $\gamma_s$ fit have very high quality. 
This is clearly seen in Figs.~\ref{fig_SagunII}, \ref{fig_SagunIII} and \ref{fig_SagunIV}.
 At the same time the results of  Ref. \cite{Becattini:gammaHIC} have typical values of  $\chi^2/dof$ between 2 and 5 at each energy point, while our value $\chi^2/dof$ = 63.4/55 = 1.15 is given for all 111 independent ratios 
 measured at 14 energies. Therefore, we conclude that the results on the strangeness suppression in heavy ion collisions reported in  \cite{Becattini:gammaHIC} are based on a low quality fit and, hence, they  cannot be regarded as the statistically confident ones.

Now we study what ratios are  improved at different energies.
For AGS energies $\sqrt{s_{NN}}$ = 2.7, 3.3, 3.8 and 4.3 GeV the description quality is quite  good even within the ideal gas model \cite{KABAndronic:05} since the number of model parameters equal or almost equal to the 
available number of  ratios and only kaons and $\Lambda$ contain strange quarks. Our detailed analysis demonstrate the fit instability for the low energies  due to two local minima  existence with very close $\chi^2$ values. Thus, for $\sqrt{s_{NN}}=$  3.8 GeV  we obtained
 $\gamma_s \simeq 1.6$ for the deepest minimum,  while   for  another  minimum, next to the deepest one,   $\gamma_s \simeq 0.8$. 
An existence of two local minima with  close values of $\chi^2$  at $\sqrt{s_{NN}} = 2.7-4.3$ GeV   allows  us to conclude that the $\gamma_s$ concept has to be refined  further in order to resolve this problem.

For the collision energy $\sqrt{s_{NN}} =$ 4.9 GeV  there are no sizeable improvements compared to  the $\gamma_s$=1 approach \cite{KABugaev:Horn2013}. The most significant data fit improvements are shown  in Figs. \ref{fig_SagunII}, \ref{fig_SagunIII}.  At the energies $\sqrt{s_{NN}} =$ 6.3 - 12 GeV the  $K^+/\pi^+$ ratio is notably improved, while the description of other ratios was  improved only slightly or even got worse. The typical examples of such a behavior are shown   in the upper and middle  panels of Fig.~\ref{fig_SagunIII}. At the same time  for collision energy $\sqrt{s_{NN}} =$ 130 GeV we find the opposite data description behavior.  The lower panel of Fig. \ref{fig_SagunIII}
  clearly demonstrates a fit quality improvement for  all ten ratios except the $K^+/\pi^+$ ratio.

An  important result of the $\gamma_s$ fitting approach is a precise  Strangeness Horn description with $\chi^2/dof = 3.3/14$, i.e. even better than it was done in \cite{KABugaev:Horn2013} with $\chi^2/dof = 7.5/14$. 
Even the highest  point of the Strangeness Horn is perfectly  described now, that makes our theoretical Strangeness Horn  as sharp as an experimental one (see upper panel on Fig. \ref{fig_SagunIV}). 

The obtained overall value  $\chi^2/dof  \simeq 1.15$ for  the $\gamma_s$ fit is only slightly better compared to the result $\chi^2/dof  \simeq 1.16$ found in   \cite{KABugaev:Horn2013}. Moreover,  the  $\gamma_s$ fit does not essentially improve either the ratios with the multistrange baryons  or the  $\Lambda/ \pi^-$ ratio (see the lower panel on Fig. \ref{fig_SagunIV}) which is a consequence of the $\bar \Lambda$ anomaly reported in \cite{AGS_L3, KABugaev:Stachel}.  Therefore,  we believe that further  improvement of the data description  is possible.


\section{Conclusions} \label{Conclusions}

We  present  an advanced description of the experimental hadron multiplicity ratios measured at AGS, SPS and RHIC energies. The inclusion of the $\gamma_{s}$ factor into the recently developed  version of the HRGM   with  the multicomponent hard-core repulsion has essentially  improved the Strangeness Horn description to $\chi^2/dof=3.3/14$, i.e.  better than it was done recently in \cite{KABugaev:Horn2013} with $\chi^2/dof=7.5/14$ and much better than it was done  in \cite{Thermal_model_review, KABAndronic:05, Becattini:gammaHIC,PBM:gamma}. For the first time even the highest point of the  Strangeness Horn is perfectly reproduced by our HRGM, which  makes our theoretical horn as sharp as an experimental one. 
In contrast to earlier results reported in  \cite{Becattini:gammaHIC},   we find  that  in heavy ion collisions there is a sizable enhancement of strangeness  at low collision energies  with  $\gamma_s \simeq 1.2-1.6$. 
The achieved   high quality fit of hadronic multiplicities   with   $\chi^2/dof \simeq  63.5/55 \simeq$ 1.15 gives us a high confidence in our   conclusions. However, 
the present   analysis shows that the $\gamma_{s}$  fit does not sizably  improve the description of the multi-strange baryons and antibaryons. Therefore, we conclude that the alternative approaches to the chemical FO suggested in \cite{BugaevEPL:13,Bugaev_SFO_13} should be developed further. We hope that  the high quality data expected to be measured at the future heavy ion facilities will help us to understand the reason of the apparent chemical 
non-equilibrium of  strange charge.

\vskip3mm 
\noindent
{\bf Acknowledgments.} We would like to thank A. Andronic for  providing an access to
well-structured experimental data.
K.A.B., D.R.O., A.I.I. and V.V.S.   acknowledge  a partial  support of the
Program `On Perspective Fundamental Research in High Energy and Nuclear
Physics' launched by the Section of Nuclear Physics  of NAS of Ukraine.
Also K.A.B.  and I.N.M. acknowledge  a partial support provided by the Helmholtz
International Center for FAIR within the framework of the LOEWE
program launched by the State of Hesse.


\end{document}